\documentclass[]{aastex631}
\usepackage{amsmath}
\usepackage{pifont}

\newcommand{\eb}{\begin{equation}}
\newcommand{\ee}{\end{equation}}

\definecolor{rkka}{RGB}{180,12,15}
\definecolor{nsgreen}{rgb}{0.1,0.5,0.1}

\shorttitle{Distributions of wide stellar binary orbits}
\shortauthors{Makarov}

\begin{document}

\title{Distributions of wide binary stars in theory and in Gaia data:\\
III. Orbital momenta, masses, and manifestations of MOND}

\correspondingauthor{Valeri V. Makarov}
\email{valeri.makarov@gmail.com}

\author[0000-0003-2336-7887]{Valeri V. Makarov}
\affiliation{U.S. Naval Observatory, 3450 Massachusetts Ave NW, Washington, DC 20392-5420, USA}

\begin{abstract}
Using the censored catalog of 103169 resolved Gaia DR3 binary stars with accurate astrometric data for each component, a new observable, object-specific parameter is computed for each pair: the projected orbital momentum. This parameter is the product of four functions of physical characteristics: total mass, semimajor axis, eccentricity, and inclination angle. Using the previously estimated marginal probability densities of eccentricity and semimajor axis, and assuming an isotropic orientation of binary systems, the sample distribution of mass was adjusted using a concordance metric of the observed and synthetic distributions of orbital momenta and an ad hoc functional model. The best-fitting mass density model is found to faithfully reproduce the observed dependence of orbital momenta on apparent separation, although the absolute luminosity distributions indicate a tendency of the widest systems to more frequently include solar-type primaries. The anticipated manifestation of MOND is computed in the investigated parameter space \{separation, momentum\}. This effect is absent in the given data. The median total mass of the widest Gaia binaries is found to be somewhat higher than that of the tighter pairs, which is interpreted as a dynamical age effect.

 \end{abstract}

\section{Introduction} \label{int.sec}
Estimating physical characteristics of wide resolved binary stars is a daunting task, due to their slow apparent motion and a deficit of observational constraints. A relative orbit (secondary with respect to primary) of a binary system is defined by seven orbital elements, while we have only four astrometric values (relative velocity and separation components) at our disposal from the Gaia mission data. For a fraction of the identified pairs, the components' radial velocities are available, but they are of relatively low utility, due to the limited measurement precision. For pairs with orbital periods in the range from several decades to a century, two astrometric catalogs with separate mean epochs (e.g., Hipparcos and Gaia) can be used to derive important characteristics, such as the mass ratio \citep{2021AJ....162..260M}. The wider systems are of prime interest for testing the models of gravitation in the weak acceleration regime, origin of exotic sources of radiation, cosmogony of stellar systems, etc. \citep{2025CoSka..55c..21B}.

Useful statistical inference can still be obtained about the underlying population distributions of basic physical parameters by capitalizing on the high astrometric precision of the Gaia data, the large number of available systems, and certain reasonable approximations and assumptions. Instead of trying to characterize each individual system, we can deduce and sometimes test specific models of how the parameters of interest are distributed and whether they are statistically independent. This approach radically reduces the degrees of freedom leading to definitive conclusions about the general properties of the sample. For example, it is not possible to determine the orbital phase (i.e., the mean anomaly) of a wide pair, but it is logical and reasonable to assume that it is uniformly distributed in the given sample. We should start with simpler models, such as a product copula for the probability density of mass and eccentricity, or Newtonian gravity versus MOND \citep{1983ApJ...270..365M}, and move to more complex alternatives only if these starting schemes fail to represent the measurements.

The statistical approach greatly benefits from derivative (or secondary) parameters computed from the available data. For example, the relative projected velocity $\Delta v$ of the components, computed from the proper motions and parallax, is sensitive to the total mass and eccentricity \citep{2004RMxAC..21....7T, 2005AJ....129.2420M, 2020MNRAS.496..987T}. The angle between the separation and velocity vectors is a separate (but not independent) secondary parameter that provides an additional constraint for Monte Carlo population synthesis \citep{1998AstL...24..178T, 2016MNRAS.456.2070T}. In this paper, we introduce a new secondary parameter $\nu$, which is the projected orbital momentum. It is directly computed from the Gaia-determined proper motions and parallax for each specific pair, combining the velocity magnitude and the motion angle.

The goal of this study is to investigate the shape and dispersion of the empirical orbital momentum as a function of separation or semimajor axis and to determine if the sample distribution can be reproduced in Monte Carlo simulations using the previously estimated sample probabilities of eccentricities and semimajor axes. The additional physical parameter entering this analysis is the total mass of the systems. It will be demonstrated that a functional model of the sample distribution of mass can be adjusted iteratively in such a way that the simulated distribution of  $\nu$ closely matches the observed histogram. The marginal probability density of mass allows us to elucidate two outstanding issues: 
\begin{enumerate}
    \item Is there evidence in the Gaia binary data that the total mass is statistically correlated with the orbit size?
    \item Can we find any manifestations of MOND in the motion of the widest pairs?
\end{enumerate}

\section{The cleaned data sample}
\label{data.sec}
The data sample used in this study is identical to the collection of resolved binary systems explored in \citep{2025AJ....169..113M, 2025AJ....170..138M}. The large, all-sky catalog of candidate binary systems (1.3 million pairs) by \citet{2021MNRAS.506.2269E} served as the primary source of information. That catalog is based on the data from the third Gaia mission release \citep{2016A&A...595A...1G, 2023A&A...674A...1G}. I have applied additional filters designed to remove the possible contaminants such as chance alignments of field stars and members of stellar clusters and associations. A stringent cut was made at the cataloged probability of chance alignment ${\cal R}<0.01$ (the authors' recommended threshold value is 0.1). Statistical interlopers of this type are especially dangerous for the proposed test of alternative gravitation using the widest pairs, because the rate of chance alignments is expected to increase as the square of separation. All distant pairs with the mean parallax below 4 mas were removed from the working sample because the astrometric measurements of distant pairs are more sensitive to the effects of unresolved inner subsystems \citep{2021A&A...649A...5F}. The original pairs with angular separations $<2\arcsec$ were also discarded, because the quality of astrometric data may be impaired by the limited angular resolution of Gaia. This filter is of low significance for the  present study, which is primarily focused on separations $a>7$ KAU. Another stringent filter on the Gaia-specific {\tt ruwe} parameter was designed to remove remaining hierarchical multiple systems with unresolved inner binaries. The applied cut was {\tt ruwe}$<1.3$ for both components.

The filtered sample was drastically reduced from an initial 1.3 million to 103,169. The number statistics in this case was traded for superior data quality and reliability. The range of projected separation $s$ in the working sample is from 25.04 AU to 161.5 KAU, with only 1\% of the sample shorter than 193.8 AU. The obvious dearth of close binaries in the source catalog is caused by the limited angular resolution of Gaia astrometry \citep{2021A&A...649A...5F}. We find 1\% of the filtered sample to be above 24461 AU, which confirms the presence of pairs with semimajor axes $a>12$ KAU, since $a$ cannot be smaller than $s/2$.

\section{Specific orbital momentum} \label{mom.sec}
The input catalog of binary systems includes individual equatorial coordinates $\{\alpha,\delta\}$ in the ICRS, proper motions $\{\mu_{\alpha}\,\cos \delta,\mu_\delta\}$, and parallaxes $\varpi$ for each component. It is straightforward to compute from these data the relative (secondary with respect to primary) position offsets $\{x,y\}$ and angular velocities $\{\dot x,\dot y\}$ in the local tangential plane associated with the primary position. The $x$ and $y$ components correspond to the tangent east and north directions, respectively. The units of choice are mas and mas yr$^{-1}$, respectively. The angular velocity vector can be further transformed into the relative tangential velocity $\{v_x,v_y\}$ in km s$^{-1}$ by using the well-known scaling relation $v=4.74\,\mu/\varpi$. Similarly, the angular offset is converted into a physical separation $\{s_x,s_y\}$ in AU by division by parallax. The novel parameter explored in this study is the vector product of the projected offset vector $s$ and the projected relative velocity $v$:
\eb 
\nu = s_x\,v_y - s_y\,v_x.
\label{nu.eq}
\ee 
It is easily verified by using the orbital equations in the tangential plane via, for example, Thiele-Innes parameters \citep{1978GAM....15.....H} that this construct is the projected specific orbital momentum:
\eb 
\nu=\sqrt{G\,M_{\rm tot}\,a}\,\beta\,\cos i,
\label{eq.eq}
\ee 
where $G$ is the constant of gravitation, $M_{\rm tot}=M_1+M_2$ is the total mass, $\beta=\sqrt{1-e^2}$, $e$ is the eccentricity, and $i$ is the angle between the line of sight and the orbit's normal, i.e., inclination of the orbit to the tangent plane. It is convenient to express mass in $M_{\sun}$, $a$ and $s$ in AU, velocity in km s$^{-1}$, and $\nu$ in km AU s$^{-1}$, in which case the gravitational constant $G=887.35$ km$^2$ AU s$^{-2}$ $M_{\sun}^{-1}$, and its square root provides the benchmark value of the Earth-Sun specific orbital momentum (29.788 km AU s$^{-1}$).

We note important assets of this parameter: it is object-specific and derived directly from the available astrometric measurements via Eq. \ref{nu.eq}, being independent of the three unknown orbital elements (two Euler angles and orbital phase), owing to the conservation of angular momentum in isolated systems. One could compute any of the parameters involved in Eq. \ref{eq.eq} (mass, eccentricity, semimajor axis, and inclination) for each pair directly from observations, if the other three were known. In reality, none of the four parameters is known for a given pair, and Eq. \ref{eq.eq} can only be used for statistical inference from large samples of binaries. 

\begin{figure*}
    \includegraphics[width=0.47 \textwidth]{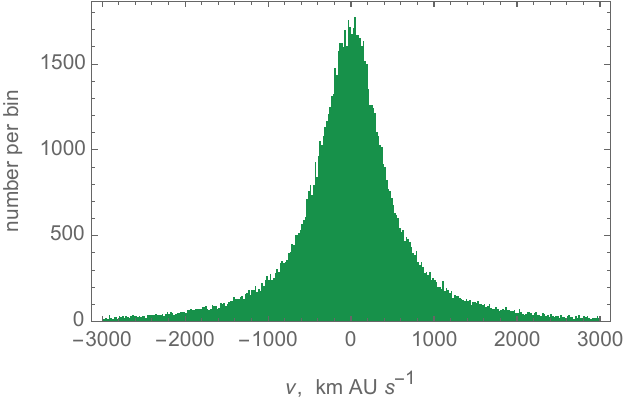}
    \caption{The histogram of projected specific orbital momenta of 103,169 binary systems.} 
    \label{hist.fig}
\end{figure*}

The histogram of the projected momenta $\nu$ for the censored sample of 103,169 pairs of stars is shown in Fig. \ref{hist.fig}. The values on $\nu$ have sign, which only reflects the sense of orbital motion in the sky plane (positive for clockwise and negative for counterclockwise, as seen on the sky). The histogram is quite symmetric around zero, where its core peaks, and has extended and shallow tails stretching to $>5000$ km AU s$^{-1}$. As expected, there is no asymmetry in the sense of rotation for this sample. For this study, the important information is contained in the shoulders of the distribution, while the core is relatively inconsequential. This is seen from Eq. \ref{eq.eq}, where $\nu$ is proportional to $\sqrt{a}$.  The widest binaries, where we can hope to find deviations from Newtonian gravity or a different distribution of masses, should have high values of $\nu$. 

The general sample sorted by separation $s$ was divided into 18 equal-sized partitions, and a $\nu$ histogram was investigated for each partition. This experiment confirmed that the core of the general sample distribution predominantly includes pairs with small values of $s$ (statistically associated with small $a$). As the average separation increases in the progression of $s$ partitions, the $\nu$ histograms become flatter and more extended. I found that the majority of the partition histograms are well-represented by the Logistic distribution PDF\footnote{{url https://mathworld.wolfram.com/LogisticDistribution.html}} with a nearly zero mean and a fitted scale parameter, which steadily grows from 125 km AU s$^{-1}$  for $s\simeq 270$ AU to 2830 km AU s$^{-1}$ for $s\simeq 50$ KAU. This confirms that the widest pairs populate the outstretching tails of the histogram in Fig. \ref{hist.fig}.

\section{Median orbital momentum versus separation}
\label{moms.sec}
It is convenient to rewrite Eq. \ref{eq.eq} in the (decimal) logarithmic form
\eb 
\log{|\nu|} = \frac{1}{2} \log{(G\,M_{\rm tot})} +\frac{1}{2} \log{a} +\log{\beta}+\log{|\cos{i}|}.
\label{lof.eq}
\ee 
We note that the $|\cos i|$ term is obviously independent of the other variables because of the isotropic distribution of orbit axes in space, and that the probability density (PDF) of $|\cos i|$ is a unity on the support $[0,1]$. In the previous analyses of the sample, statistical independence of the distributions of $e$ (or $\beta$) and $a$ has been assumed for the population of wide pairs that dominates the sample, and no evidence emerged to the contrary. \citet{2020MNRAS.496..987T}, using a different method, estimated the sample distribution of eccentricity for an earlier Gaia dataset and detected a dichotomy of eccentricity probability density functions (PDF) at different parts of the apparent separation range. While closer pairs with separations $<200$ AU have smaller eccentricities than the sample average, the wider systems at $s>1$ KAU tend to have eccentricities closer to 1, and their PDF can be approximated as $e^{1.2}$. Based on these previous studies, we model the marginal PDF of eccentricity with the power-law function:
\eb 
P_e(e)= (2-\alpha')e(1-e^2)^{-\alpha'/2},
\label{pe.eq}
\ee 
For the sample under investigation, the best-fitting power index $\alpha'=+0.15$ was determined using the motion angle observable parameter, which is especially sensitive to the distribution of eccentricity but independent of mass or total energy \citep{2025AJ....169..113M}. This distribution is slightly concave with a rise toward $e=1$. It is functionally very close to the PDF$[e]=11/5\,e^{6/5}$ estimated by \citet{2020MNRAS.496..987T}. No evidence was found that $\alpha'$ may be dependent on the orbit size at $a>1$ KAU. 

The numerically reconstructed distribution of semimajor axis $a$ was found to be well approximated with a power-law relation,
\eb 
P_a(a)=K\, a^{-0.127-0.414\,(\log a - 3)}
\ee 
for $\log{a}>3$ \citep{2025AJ....170..138M}. In the $\log$--$\log$ space, this relation follows a straight declining line at least up to $\log{a}\simeq 4.5$. Alternative functional forms for $P_a(a)$ were also proposed, based on shifted Gamma distributions. Unfortunately, these functional models cannot be used in this study, for the following reasons. The distribution of $a$ cannot be marginalized in any MC simulations involving the observed relative projected velocity or orbital momentum. The relative velocity is inversely proportional to $\sqrt{a}$, whereas the specific momentum is proportional to $\sqrt{a}$. Therefore, the given vector of separations $s$ should be coupled with the given vector of kinematic parameters. It would be incorrect to assume that all pairs with the same $s$ have statistically the same $a$, because the mapping between these parameters is strongly asymmetric. Therefore, we have to step back and implement the statistical mapping $s\rightarrow a$ for each pair in the sample using the intermediary scaled projection parameter,
\eb 
f=\frac{s-a}{a},
\label{f.eq}
\ee 
which can be MC-simulated in a separate procedure. This parameter is independent of the physical parameters $a$ and $M_{\rm tot}$. It is dependent on the PDF($e$) (or, within the adopted model, on $\alpha'$), but this dependence was found to be rather weak.

\section{Implementation of MC simulations} \label{imp.sec}
To estimate the expected relation between the projected momentum $\nu$ and projected separation $s$ in the Newtonian regime, a full-scale MC simulation of binary orbits was performed for a sample of $k=4\times 103169$ pairs, i.e., four times the size of the data sample. The simulation cycle begins with random number generation of $k$ normalized projection factors $f$ (Eq. \ref{f.eq}) using the separately evaluated empirical distribution of this parameter for $\alpha'=+0.15$. The fourfold vector of observed separations $s$ is converted to a random realization of the corresponding vector $a$ by dividing $s$ by $f+1$. Random vectors $\omega$, $\Omega$ of the same size are generated from uniform distributions on their supports, while a vector of inclination $i$ is drawn from the isotropic distribution
\eb 
P_i(i)=\frac{1}{2}\sin i/2.
\ee 
A random vector of eccentricity $e$ is generated from the distribution Eq. \ref{pe.eq}, which is internally consistent with the realization of $f$. To this compendium of observational and synthetic data, a random vector of mean anomaly instances (uniformly distributed on $[0,2\pi[$) is added; it can be converted to eccentric anomalies $E$ using, for example, the back-interpolation scheme \citep{2018arXiv181202273T}. This information is sufficient to compute $k$ sets of four Thiele-Innes parameters. From these, the synthetic projected separation vectors are directly computed. The entire simulation loop is validated by the close agreement of the sample distributions of the measured and synthesized $\nu$ for the high-momentum pairs, as discussed in Section \ref{res.sec}.

The remaining step is to compute the projected relative velocity vectors from the synthesized Thiele-Innes parameters and generate the noise component. This requires one additional orbital element, namely, the mean motion $n$, which should be consistent with the generated $a$ via the third Kepler law. The projected separations in AU and projected velocity components in km s$^{-1}$ are directly computed from the synthetic orbital parameters and measured parallaxes using Eqs. A1, A2, A9, and A10 in \citep{2005AJ....129.2420M}.

\section{Total mass models} \label{mass.sec}
Two ad hoc probability density models for $M_{\rm tot}$ have been tested in this study, which produced close goodness-of-fit (GoF) results:
\begin{eqnarray}
    P(M_{\rm tot})&=&{\cal G}[h,w](x-x_0) \label{Gamma.eq}\\ 
    P(M_{\rm tot})&=&{\cal R}[z](x-x_0)
\end{eqnarray}
where ${\cal G}$ stands for the Gamma distribution, ${\cal R}$ for the Rayleigh distribution, and $h$ is the shape parameter, $w$ and $z$ are the scale parameters, $x=M_{\rm tot}/M_{\sun}$, $x_0$ is a free offset parameter. These free parameters were adjusted in a nested iterative optimization based on the L1F GoF metric introduced in \citet{2025AJ....169..113M}. In this application, the L1F parameter quantifies the closeness of the MC-simulated CDF of $\nu$ to the empirical CDF$(\nu)$ across the entire range of orbital momenta. Massive MC trials for $4\times10^5$ simulated pairs were produced by taking one of fitting parameters from a grid of values (e.g., $w$ in Eq. \ref{Gamma.eq}) and fixing the other fitting parameters (e.g., $h$ and $x_0$) at the previously adjusted values. Essentially, this procedure amounts to an iterative 3D grid optimization. The updated best-fitting parameter corresponds to the globally smallest L1F value. With the updated distribution parameter fixed, the MC grid optimization was performed for another fitting parameter, etc. A few upper-level iterations proved sufficient to converge to a set of optimal fitting values.

\begin{figure*}
    \includegraphics[width=0.47 \textwidth]{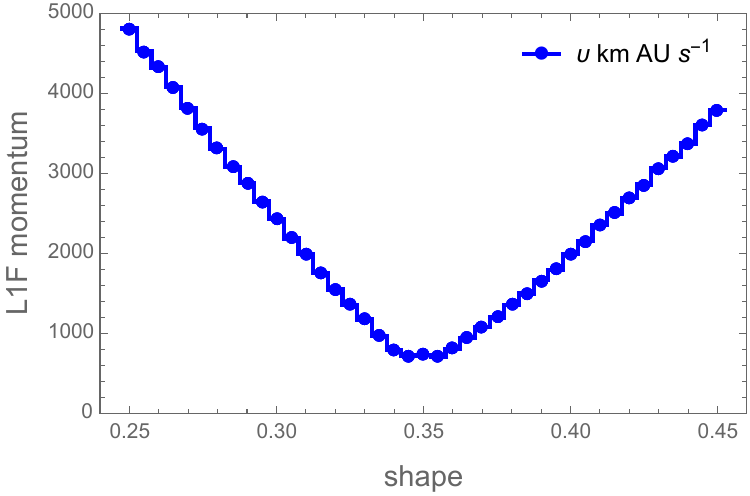}
    \caption{Distribution concordance metric L1F of Monte Carlo random samples of orbital configurations with the empirical sample from Gaia GR3 data in the projected orbital momentum parameter $\nu$ versus the assumed shape $h$ of the total mass modeled with the Gamma distribution in Eq. \ref{Gamma.eq}. The small L1F value indicates the best match of the distributions.} 
    \label{shape.fig}
\end{figure*}

Fig. \ref{shape.fig} shows the results of one such iteration for the Gamma-distribution shape parameter $h$. The L1F statistics were computed for a grid of $h$ with a step of 0.01 and $4\times10^5$ synthetic pairs. 

\begin{figure*}
    \includegraphics[width=0.47 \textwidth]{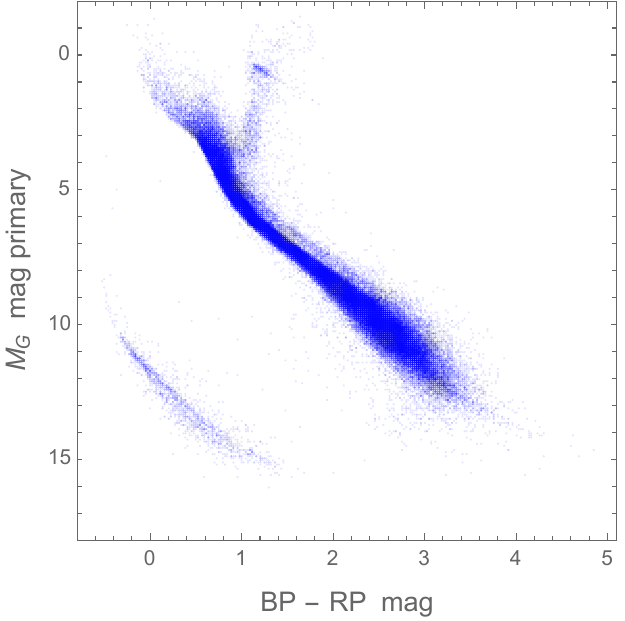}
    \includegraphics[width=0.47 \textwidth]{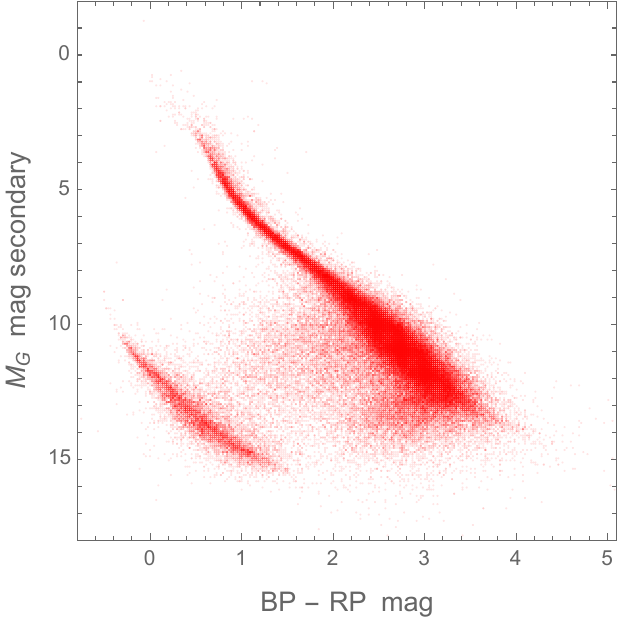}
   \caption{HR diagrams of absolute $G$ magnitudes versus BP$-$RP colors for primary and secondary components of the investigated sample of resolved binaries in Gaia DR3.}
    \label{HR.fig}
\end{figure*}

The difference between the two alternative mass distributions is that the ${\cal G}$-model has a higher skewness (1.20 against 0.63 for the Rayleigh fit) and slightly lower mean, median, and peak values. The median total mass for the ${\cal G}$-model fit is 0.85 $M_{\sun}$. It may be surprising that the typical Gaia binary has a subsolar total mass, but this result is consistent with both the relative motion data \citep{2022AJ....164..164G} and the available information about the luminosity of the resolved components. Fig. \ref{HR.fig} shows the Hertzsprung-Russell diagrams (absolute $G$ magnitude, denoted as $M_G$, versus BP$-$RP color) for the nominal primaries (left panel) and secondaries (right panel) in the same axes. The binary systems are mostly composed of main-sequence (MS) dwarfs, with a small contribution from white dwarfs (WD) and giants. A small fraction of the components are unresolved photometric binaries---these are seen as the ``shadow" main sequence shifted up and to the right by up to 0.75 mag \citep{2018A&A...616A..10G} (for a detailed description of the photocenter effect in unresolved binaries, cf. Fig. 3 in \citet{2007ApJS..173..137G}). A finite admixture of unresolved MS-WD binaries among the secondaries is also evident as a plume extending from the lower part of MS toward the dual WD sequence. Using the estimated relation between the $M_G$ magnitude and mass \citep{2023A&A...678A..19C}, quantiles of the component mass distributions can be evaluated. We find that 10\% of the 103,169 primaries in the working sample are more massive than $1.4\,M_{\sun}$, 10\% are smaller than $0.3\,M_{\sun}$, and the median mass is about $0.58\,M_{\sun}$. The corresponding values for the secondary components are $0.58\,M_{\sun}$, $0.15\,M_{\sun}$, and $0.3\,M_{\sun}$. The sum of the median component masses seems to be consistent with the median of the modeled total mass. The photometric total mass may be slightly underestimated due to the presence of white dwarfs and unresolved multiples.

\section{Generation of random noise in orbital momenta}
In the context of testing the presence of excessive relative motion in the widest binary systems, a realistic modeling of observational errors is crucial. Eq. \ref{nu.eq} shows that the input noise comes into the determination of $\nu$ via both the components of projected separation $s$ and components of projected velocity $v$. The errors of $s_x$ and $s_y$ originate from the measurement errors of angular positions and the parallax error via the conversion of mas to AU. The former are negligibly small compared with the other contributors to the error budget. The parallax error propagates into $\nu$ estimate in a nontrivial way via both $s$ and $v$ input parameters. Distant binaries with small true parallaxes also introduce the well-known sampling bias \citep{1973PASP...85..573L}, which makes the estimated velocities and separations systematically larger than the true values. This can have a catastrophic effect for the widest pairs with low signal-to-noise ratios in relative proper motions, and the parallax filter $\varpi>4$ mas implemented here is meant to eliminate the bias.

Even with the implemented filter on parallax, observational noise remains a major factor in computation of projected orbital momenta through the propagation of proper motion errors. A significant fraction of pairs have rather low signal-to-noise ratio determinations because of this factor. False conclusions can be drawn about the presence of excess momentum in the widest pairs unless this component is carefully incorporated in the analysis. Here, this  task is performed in the most direct and statistically robust fashion by generating random noise perturbations of the observable parameters. For each observed pair of stars, four synthetic velocity vectors are generated with randomly chosen masses and semimajor axes (consistent with the observed separation), as described in Section \ref{imp.sec}. A random perturbation vector $\boldsymbol{\epsilon}$ is added to each of these realizations, which is computed from a random vector of proper motion error $\boldsymbol{\mu}$:
\eb 
\boldsymbol{\epsilon}=4.74\, \boldsymbol{\mu}/\varpi,
\ee 
and $\boldsymbol{\mu}$ is drawn from the binormal distribution \citep{Binormal} ${\cal N}_2[\boldsymbol{0},\boldsymbol{\Sigma}_\mu]$, where $\boldsymbol{\Sigma}_\mu$ is the given $2\times2$ proper motion block of the formal covariance matrix. At this point, the specified covariance $\boldsymbol{\Sigma}_\mu$ can be manipulated by adding a fixed noise floor to the variances in the diagonal and then multiplying the entire matrix by a fixed noise scale ($\geq 1$). Such modifications are often used in geodetic VLBI astrometry, for example, where the formal errors are known to underestimate the apparent dispersion of measurements. This option was exercised in the present study, but found unnecessary because the provided formal errors for Gaia binaries seem to adequately capture the error budget.

The thus-perturbed vector $\boldsymbol{v}_{\rm syn}$ still does not fully incorporate the main sources of error, because the parallax measurement $\varpi$ is uncertain too. This is corrected by introducing a noise factor
\eb 
z=(\varpi/(\varpi+w))^2, \label{z.eq}
\ee 
where $w$ is a random scalar drawn from ${\cal N}_1[0,\sigma_\varpi^2]$, and $\sigma_\varpi$ is the given formal error of parallax. The synthetic momentum $\nu$ computed from the generated components $s$ and $v_{\rm syn}$ by Eq. \ref{nu.eq} is multiplied by $z$. Note that the square power in Eq. \ref{z.eq} takes into account the double effect of the the error in parallax when it propagates to $\nu$ via $s$ and $v$.

\section{Results}
\label{res.sec}
\begin{figure*}
    \includegraphics[width=0.47 \textwidth]{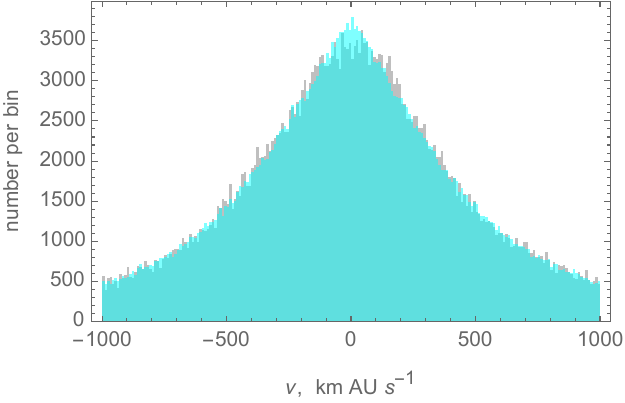}
    \caption{The histograms of observed projected specific momenta (light gray) and simulated projected specific orbital momenta (cyan) of Gaia DR3 binary stars.} 
    \label{nusimu.fig}
\end{figure*}

Before any conclusions can be drawn from the elaborate simulation scheme with multiple model distributions of basic parameters, it should be validated by a test of consistency. The most important test is how faithfully the empirical sample distribution of orbital momenta (Fig. \ref{hist.fig}) is reproduced by the synthesized data. The L1F consistency metric has already been used to adjust the adopted model of total mass distribution in Section \ref{mass.sec}. Fig. \ref{shape.fig} shows that the achieved concordance is close but not ideal. Fig. \ref{nusimu.fig} 
displays a superposition of the empirical and simulated $\nu$ distributions (in one of the trials). We can see that the wings of the distribution are reproduced extremely well. The reason for the imperfect concordance also becomes obvious. The simulated histogram does not reproduce the clipped top of the sample distribution in the close vicinity of 0. Since the total number of pairs is the same in the two histograms, the missing pairs at the peak are rather uniformly redistributed in the shoulders. The deficit of nearly zero projected momenta is probably caused by the limited angular resolution (fixed at $2\arcsec$), which effectively removes some of the closer and more distant pairs.

\begin{figure*}
    \includegraphics[width=0.47 \textwidth]{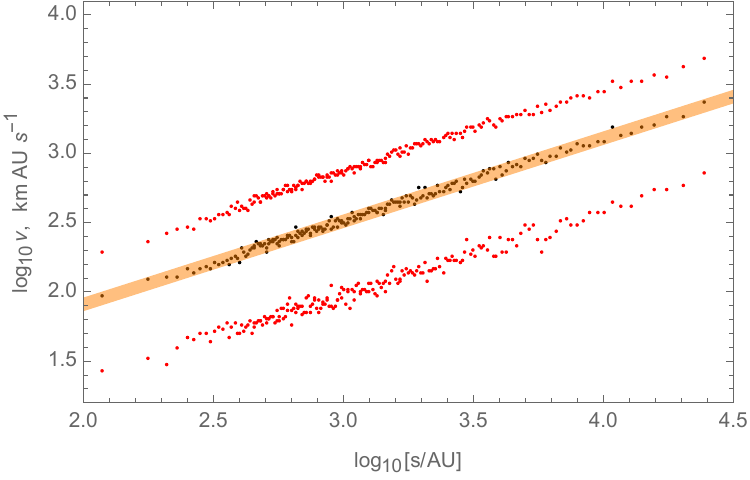}
    \includegraphics[width=0.47 \textwidth]{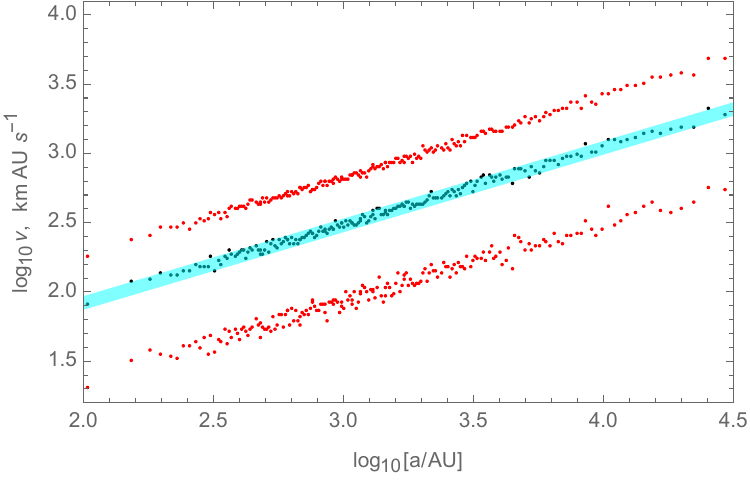}
    \caption{Binned median logarithm of orbital momentum $\nu$ (black dots) and the $\{0.16,0.84\} $ quantiles (red dots) versus the (left panel) the logarithm of observed projected separation $s$ in AU and
    (right panel) the logarithm of simulated physical semimajor axis $a$ in AU. The yellow and cyan thick lines are the fits presented in Eqs. \ref{1fit.eq} and \ref{2fit.eq}, respectively.}
    \label{nusa.fig}
\end{figure*}

Having passed the general concordance test, we can now more specifically investigate the relation between the orbit size and projected orbital momentum. Fig. \ref{nusa.fig}, left panel, shows the empirical dependence of the binned median $\nu$ versus binned median $s$ in logarithmic scales. These data were computed by sorting the given sequence of $\{s,\nu\}$ tuples by $s$, dividing the sorted sequence into 196 equal partitions of 525 pairs, and taking the median values the logarithms of $s$ and $\nu$ within each bin. The 0.16 and 0.84 quantiles of the $\nu$ values were also computed for each bin representing the empirical dispersion of individual data points within each bin. The leading portion of the data plotted in Fig. \ref{nusa.fig}, left, is reproduced in Table \ref{196.tab}, and the entire table is published online. We find that the median momenta and their quantiles follow straight lines in the logarithmic scale remarkably well in the wide range of $\log s$ between 2 and $>4.5$. The thick orange line represents the numerically optimized linear fit (1-norm nonlinear optimization without weights) of the median:
\eb 
\log{\hat{|\nu|}}=0.60\,\log[s]+0.71. \label{1fit.eq}
\ee 
The linear dependence of $\log{\hat{|\nu|}}$ on $\log a$ could be expected from Eq. \ref{lof.eq}, but the slope is certainly larger than 0.5. Note, however, that the median of the sum of independent random variables is not equal to the sum of their medians. Specifically, the skewness of the PDFs of $a$ and $M_{\rm tot}$ alter the slope of the empirical and synthesized dependencies. The steeper slope is a mathematical property, not an excess signal.

\textcolor{rkka}{
\begin{deluxetable*}{l|CCC} 
\caption{Median and quantile values of $\log_{10}(|\nu|)$ for nonoverlapping bins of sorted $\log_{10}(s/{\rm AU})$.}
\tablehead{ \colhead{$\log_{10}(s/{\rm AU})$} & \multicolumn{3}{c}{$\log_{10}(|\nu|)$}\\
\colhead{} & 
\colhead{median} & \colhead{0.16-quantile} & \colhead{0.84-quantile}  }
\decimals
\startdata
 2.07476 & 1.96964 & 1.42875 & 2.28311 \\
 2.24817 & 2.08974 & 1.52421 & 2.36916 \\
 2.3195 & 2.10889 & 1.47681 & 2.42628 \\
 2.36438 & 2.11018 & 1.58895 & 2.45481 \\
 2.39818 & 2.16441 & 1.67636 & 2.46545 \\
 2.42747 & 2.1347 & 1.65112 & 2.44713 \\
 2.45223 & 2.17213 & 1.69582 & 2.52722 \\
 2.47288 & 2.17618 & 1.69729 & 2.53035 \\
 2.49125 & 2.16876 & 1.64071 & 2.51535 \\
 2.50809 & 2.21939 & 1.71789 & 2.52554 \\
 \enddata
\tablecomments{
The entire table is available online.
}
\label{196.tab}
\end{deluxetable*}}
To prove this point, the MC-generated population of binaries was used to reveal the expected dependence of median $\nu$ on semimajor axis $a$. The right panel of Fig. \ref{nusa.fig} displays this dependence in the logarithmic scales, which was computed in the same way as the empirical dependence in the left panel, but for synthesized $a$ instead of $s$. Again, the black dots represent the binned medians, and the red dots are the corresponding 0.16 and 0.84 quantiles of $\nu$ values in each bin. The thick cyan line is the best fit:
\eb 
\log{\hat{|\nu|}}=0.56\,\log[a]+0.80. \label{2fit.eq}
\ee 
The MC-generated relation almost perfectly follows this linear fit in a wide dynamic range of orbits covering 2.5 orders of magnitude. The slight change of the slope (from 0.60 to 0.56) between the two fits does not reflect the uncertainty of this parameter but the different shape of the PDFs of $s$ and $a$ \citep{2025AJ....170..138M}. Upon convolution with the PDFs of other parameters ($M_{\rm tot}$, $\beta$, and $i$), the observed dependence of $\nu$ versus $s$ emerges to be slightly stronger than the estimated dependence of $\nu$ on $a$. These finds provide a valuable test of Newtonian dynamics in binary stars.

\section{A search for manifestations of MOND in the widest binaries}

The results presented in Fig. \ref{nusa.fig} were obtained in the framework of the null hypothesis, which involves strictly Newtonian gravitation and homomorphic distributions of mass and eccentricity across the range of orbital sizes. The null hypothesis should be rejected if we find compelling evidence that it does not adequately describe measurement data (per Occam's razor). This is not the case here, as the simplest null hypothesis produces consistent and conformal results. The projected orbital momentum shows almost identical and well-behaved relations to orbit sizes from 100 AU to higher than 30 KAU. In particular, there is no evidence of any excessive momenta for orbital separations above $\sim 7$ KAU where the MOND effects are expected to be manifest for solar-type binaries. But is this method sensitive enough for the test?

For testing modified gravity in wide binaries, it is important to quantify its effect for a chosen observable parameter and compare that prediction with the actual observation. The projected orbital momentum explored in this study is sensitive to possible manifestations of MOND. Numerical simulations of orbital trajectories in the MOND regime (including the presumed external field effect) have shown that the typical revolution time becomes significantly shorter than the Keplerian period with the same initial conditions \citep{2010OAJ.....3..156I}. Alternatively, to obtain a trajectory of roughly the same size as a given Kepler ellipse, a faster initial velocity is required. Even in isolated binary stars, the orbital parameters can only be considered as osculating elements, because the chaotically changing orbital loops transform the orientation and eccentricity on timescales comparable to the anomalistic ``period". In our case, the initial or mean velocity is a free parameter, while the initial separation is constrained by the data. Therefore, it is reasonable to 
introduce a modified (osculating) mean motion $n_{\rm MG}$ while fixing the given semimajor axis $a$.
It follows that the modified instantaneous orbital acceleration of the secondary with respect to the primary is
\eb 
g_{\rm MG}=a\,n_{\rm MG}^2=\frac{g_{\rm N}}{2}\left(1+\sqrt{1+4\,g_0/g_{\rm N}}\right),
\ee 
where $g_{\rm N}=G\,M_{\rm tot}/r^2$ is the Newtonian acceleration, $g_0$ is the basic MOND acceleration parameter ($g_0=1.2\times 10^{-10}$ m s$^{-2}$), and we made use of the simple interpolation formula for
$g_{\rm MG}$ \citep{2012LRR....15...10F}. This further simplifies to
\eb 
n_{\rm MG}=\frac{n}{\sqrt{2}}\left(1+\sqrt{1+4\,g_0/(a\,n^2)}\right)^{1/2},
\ee 
which can be used in MC simulations of $\nu$ from synthetic distributions of $a$ and $M_{\rm tot}$.

\begin{figure*}
    \includegraphics[width=0.47 \textwidth]{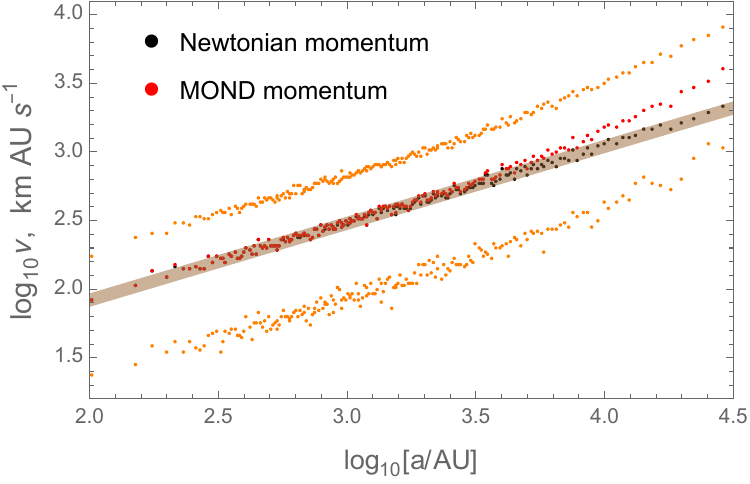}
    \includegraphics[width=0.47 \textwidth]{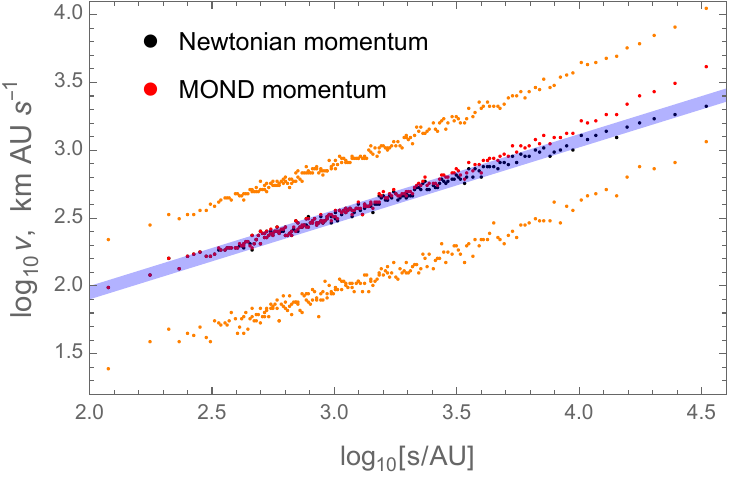}
   \caption{Binned median logarithm of orbital momentum $\nu$ (black dots) 
   as a function of simulated semimajor axis $a$ (left panel) or observed separation $s$ (rigth panel) for the working sample of Gaia binaries with two theories of gravitation: Newtonian dynamics (black dots) and MOND (red dots). The orange dots depict the corresponding $\{0.16,0.84\} $ quantiles of the MOND version of orbital momentum, which shows a clear deviation from the Newtonian prediction starting at $\log a=3.5$. The thick brown line in the left panel is the same fit as in  Eq. \ref{2fit.eq}, also shown in Fig. \ref{nusa.fig} (right panel).}
       \label{numond.fig}
\end{figure*}

Full-scale MC simulations of $\nu$ were performed in the MOND regime using the methods and models described in Sections \ref{imp.sec} and \ref{mass.sec}. The random realizations of $a$ were anchored to the actual observed separations $s$, parallaxes $\varpi$, and the formal covariances of relative proper motions and parallaxes. A realistic simulation of random noise is crucial for the pairs with wide separations. The previously explored in the literature observable parameter $v$ (relative projected velocity) has the unfortunate property of declining with separation. The widest systems are then in the low-SNR domain, where the formal uncertainties become comparable to the signal. Furthermore, the uncertainty of parallax also contributes to this estimation in a nontrivial way, because it is involved in the conversion of proper motion to velocity. Simulation of measurement noise should be performed from the bottom up, because the given formal errors of proper motion components are not copied over to the magnitude of velocity vector (recalling that the mean of the $\chi[2]$ distribution equals 1.25). These caveats justify the generic synthesis of individual measurement errors with their correlations for each pair in the inflated working sample. The results of a complete duty cycle simulation are presented in Fig. \ref{numond.fig}. The black dots show 
the $\log{\hat{|\nu|}}$ dependence on $\log[a]$ (left panel) and on $\log[s]$ (right panel) assuming the Newtonian mean motion $n$, the red dots represent the same dependencies with the MOND mean motion $n_{\rm MG}$, and the sequences of orange dots show the 0.16 and 0.84 quantiles for the MOND simulation. The brown line is the same linear fit as in Fig. \ref{nusa.fig} (right panel) and Eq. \ref{2fit.eq}. Note that the black dots in the left plot are not identical to those in Fig. \ref{nusa.fig} because it is a different MC-realization.

It is evident that the simulated momentum $\nu$, which is consistent with the observed momentum, obeys the Kepler law and does not show the excess starting around $\log a=3.5$ predicted by MOND. We thus find no reasons to discard the null hypothesis for the investigated sample.

\section{Conclusions}
\label{end.sec}

The novel parameter $\nu$ explored in this study is the magnitude of the specific orbital momentum vector projected onto the sky plane. It can be directly computed from the available Gaia data for each pair of resolved binaries. Owing to the momentum conservation law, this value depends only on four object-specific parameters: total mass, semimajor axis, eccentricity, and inclination angle (Eq. \ref{eq.eq}). It is independent of the other two orbit's orientation angles and the instantaneous anomaly. This property makes it potentially useful for estimating poorly known physical characteristics of wide binaries such as total energy and mass.

A clearly filtered sample of the most reliable 103,169 binary systems shows a strong dependence of the distribution of $\nu$ values on observed (projected) separation $s$. At a fixed $s$, this distribution is well-represented by a Logistic distribution with a certain scale parameter. The scale steadily increases with $s$, which reflects a similar dependence of the distribution width on $a$. Indeed, half of the pairs at a fixed $s$ have semimajor axes $a>1.017\,s$ for the empirically estimated eccentricity power index $\alpha'=+0.15$. The widening scale of orbital momenta with $a$ is a fully expected effect (Eq. \ref{lof.eq}).

It is tempting to use the previously estimated PDFs for $a$ and $e$ (and the a priori known distribution of $\cos i$) in a straightforward MC generation of $\nu$ to constrain the total mass distribution and test MOND. Unfortunately, these objectives require a full-scale simulations of individual orbits from bottom up (including orbit orientation angles and orbital phase) because the random measurement noise has to be carefully and accurately modeled too. The effect of measurement errors is {\it not} uniform across the range of separations. The relative velocity of the components, which is computed from the measured proper motion and parallax, is more perturbed at the high end of separations, where it can show a false apparent excess. In a separate adjustment, an ad hoc model should be fitted representing the sample PDF of total mass (not to be confused with the initial mass function). Two shifted bell-shaped models based on Gamma and Rayleigh PDFs were found to yield close concordance metrics L1F with their optimized parameters on the MC-generated samples of $\nu$. A typical binary star in the Gaia collection has a total mass well below $1\,M_{\sun}$, in agreement with the empirical color-magnitude diagram (Fig. \ref{HR.fig}). The sample is dominated by late dwarfs with a smaller admixture of white dwarfs and primary giants. 

The implemented full-scale generation of observable parameters $s$ and $\nu$ revealed excellent agreement with the data across the entire dynamic range except for the vicinity of zero orbital momentum (Fig. \ref{nusimu.fig}). There is a deficit of small momenta in the given sample. The likely reason is the limited angular resolution and the related lower bound on $a$. This feature is not important for the topic of this study, which concerns the widest separations and high orbital momenta. The simulations were also able to reproduce very well the empirical dependence of the median $\nu$ on separation (Fig. \ref{nusa.fig}). The linear fits (Eqs. \ref{1fit.eq} and \ref{2fit.eq}) are nearly perfect in the range from 100 AU to $>30$ KAU. There is no excess orbital momentum that could be attributed to MOND in the investigated sample of $10^5$ binaries.

To confirm that the proposed test is sensitive to the expected MOND effects, a separate cycle of full-scale MC simulations was performed following the same algorithms and statistical models but replacing the Newtonian acceleration with the boosted MOND acceleration while fixing the observed orbital separations. The expected signal is evident as a departure of the median $|\nu|$ from the simple power-law fit in Fig. \ref{numond.fig}. The excess in $|\nu|$ caused by MOND is up to a factor of 2 at $\log a=4.5$, which would be easily detectable. A similar excess would be seen in the median $|\nu|$ as a function of $\log s$. Previous analyses of the expected MOND-related signal were mostly based on the relative projected velocity parameter and resulted in conflicting conclusions \citep{2023OJAp....6E...2M, 2024ApJ...960..114C, 2025A&A...699A.151S, 2024MNRAS.533..729H, 2025MNRAS.537.2925H, 2024MNRAS.527.4573B, 2023OJAp....6E...4P, 2019MNRAS.488.4740P, 2025OJAp....8E.109P}. This study contributes to the accumulated evidence that the wide binaries' dynamics is adequately described by the classical theory of gravitation. A likely reason for the false positive results is inadequate treatment of the astrometric noise components. In particular, Eq. 7 in \citep{2024ApJ...960..114C} used to estimate the uncertainty of the relative proper motion magnitude is incorrect. It leads to a strongly underestimated noise component for the widest separations where the SNR is tending to zero.

\begin{figure*}
    \includegraphics[width=0.47 \textwidth]{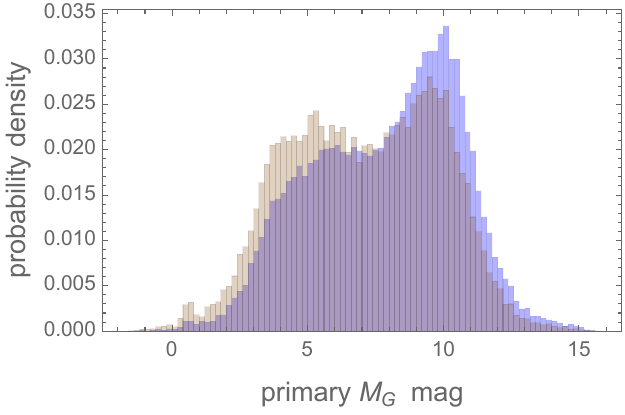}
    \includegraphics[width=0.47 \textwidth]{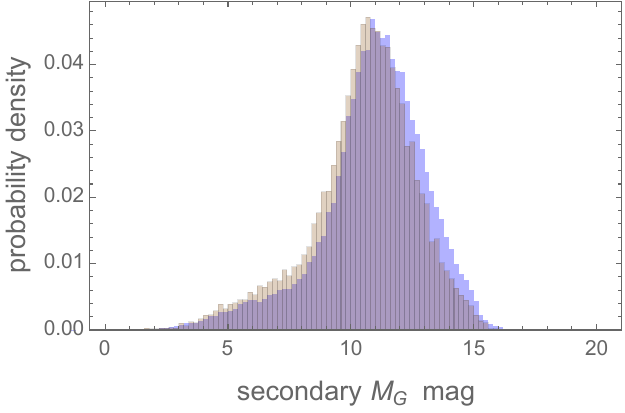}
    \caption{Histograms of absolute $G$ magnitude $M_G$ for primary (left plot) and secondary (right plot) components of resolved Gaia binaries divided into two subsets: pairs with $\log s<3.5$ (blue shade) and $\log s>3.5$ (brown shade). All histograms are normalized to unit area.}
    \label{abs.fig}
\end{figure*}

While Occam's razor dictates that we should accept the null hypothesis, the possibility of a negative result arising from the adopted models and assumptions should also be considered. Eq. \ref{eq.eq} tells us that, among the physical parameters, only $M_{\rm tot}$, $\beta$, and $i$ could possibly conspire to hide a true excess of orbital momentum at large $a$. The assumptions about $i$ and its independence of the other characteristics are solid, as we would need predominantly edge-on orbits for the wide pairs to explain the absence of signal. Likewise, the distribution of eccentricity would have to be much steeper than the previously determined power-law with $\alpha'=+0.15$, to the extent that all wide pairs are parabolic. Previous studies provided results that are generally consistent with the superthermal probability density adopted in this study at wide separations \citep{1998AstL...24..178T, 2016MNRAS.456.2070T, 2020MNRAS.496..987T, 2022MNRAS.512.3383H}.

The total mass is a new contributing factor, which was assumed to be independent and homomorphic across the range of orbit sizes (Section \ref{mass.sec}). The orbital momentum is proportional to the square root of $M_{\rm tot}$; hence, the widest pairs have to be less massive by a factor of 0.25 on average to account for the missing \textcolor{rkka}{MOND-predicted} excess at $\log s=4.5$. This supposition is testable using $M_G$ as a proxy for mass, given the fact that most components are main-sequence dwarfs (Fig. \ref{HR.fig}). The filtered sample was divided into two unequal parts: pairs with $\log s<3.5$ (subset 1) and pairs with $\log s>3.5$ (subset 2). The absolute $G$ magnitudes were compared for the two subsets and separately for the primaries and secondaries. Fig. \ref{abs.fig} shows the four probability density histograms (i.e., empirical PDFs) of $M_G$. We find that the shape and location of the $M_G$ distribution are indeed dependent on apparent separation. The closer primaries from subset 1 are distinctly less luminous (and therefore less massive) than their counterparts in wider pairs. The bump at $M_G\simeq 5$ mag (solar analogs) is much more prominent for the widest pairs. The median $M_G$ is shifted from 8.2 to 7.2 mag between the two subsets of the primaries. The same trend is observed among the secondaries with a more modest shift from 11.1 to 10.7 mag. Assuming that the mass-luminosity relation is dominated by MS dwarfs, the corresponding upward shifts of mass are $0.1\,M_{\sun}$ and $0.04\,M_{\sun}$, respectively.\footnote{From the interpolation table by E. Mamajek: {\url{https://www.pas.rochester.edu/~emamajek/EEM_dwarf_UBVIJHK_colors_Teff.txt}}} Thus, a statistical dependence of $M_{\rm tot}$ on $a$ is indeed present in the Gaia collection of binaries, but its effect is opposite to the initial supposition. The impact of the correlation is limited due to the weak dependence of $|\nu|$ on total mass, and it could only generate a spurious excess of momentum in the present analysis. The implicit $M_{\rm tot}$--$a$ correlation is interpreted as a dynamical age effect. Initial binaries with low binding energy are more easily disrupted in random encounters with unrelated stars in open clusters and in the field \citep{1975MNRAS.173..729H, 1975AJ.....80..809H, 1987ApJ...312..390W, 1987ApJ...312..367W, 2025JCAP...02..001Q}. The widest pairs appear to be more massive because they are younger, and they did not have enough time to be disrupted and to join the multitude of single stars. The closer pairs, on the other hand, are older because they are more resilient to stochastic disruption.

The absence of boosted acceleration for very wide binaries adds to the number of problems encountered in the framework of MOND, such as the violation of conservation of energy and momentum, and the non-additivity of  gravitational mass action. Although manifestly successful in explaining the flat rotation curves of spiral galaxies, MOND does not seem to pass the wide binary test. A symmetrically deep consideration should be given to alternative theories that are consistent with this dichotomy. The scalar-tensor-vector modification of gravity, in particular, predicts that the dynamics of the Solar system and wide binary stars is governed by Newton's law \citep{2024JCAP...05..079M}.

\bibliography{main}
\bibliographystyle{aasjournal}

\end{document}